\documentclass{ws-mpla}
\usepackage[super]{cite}
\usepackage{graphicx}
\begin{document}

%\markboth{Parthapratim Pradhan}

%%%%%%%%%%%%%%%%%%%%% Publisher's Area please ignore %%%%%%%%%%%%%%
\catchline{}{}{}{}{}
%%%%%%%%%%%%%%%%%%%%%%%%%%%%%%%%%%%%%%%%%%%%%%%%%%%%%%%%%%%%%%%%%%%

\title{Area Product and Mass Formula for Kerr-Newman-Taub-NUT Space-time}

\author{Parthapratim Pradhan}

\address{Department of Physics,\\ Vivekananda Satavarshiki Mahavidyalaya \\
 (Affiliated to Vidyasagar University),\\ Manikpara, Jhargram, West Midnapur,\\
  West Bengal~721513, India\\ E-mail: pppradhan77@gmail.com }

\maketitle

\pub{Received (Day Month Year)}{Revised (Day Month Year)}

\begin{abstract}
We derive area product, entropy product, area sum and entropy sum of the event horizon and Cauchy horizons for
Kerr-Newman-Taub-NUT(Newman-Unti-Tamburino) black hole  in four dimensional \emph{Lorentzian geometry}. We observe 
that these thermodynamic products are \emph{not} universal(mass-independence) for this  black hole(BH), whereas for 
Kerr-Newman(KN) BH such products are universal (mass-independence). We also examine the entropy sum and area sum. It is shown
that they all are depends on mass, charge and NUT parameter of the back ground space-time. Thus we can conclude
that the universal(mass-independence) behaviour  of area product and entropy product,  area sum and entropy sum for
Kerr-Newman-Taub-NUT(KNTN) BH fails and which is also quite different from KN BH. We further show 
that the  KNTN BH do not possess \emph{first law of BH thermodynamics } and  \emph {Smarr-Gibbs-Duhem } relations,
and that such relations are unlikely in the KN case.  The failure of these aforementioned features are due to presence 
of the non-trivial NUT charge which makes the space-time to be asymptotically non-flat, in contrast with KN BH. The another
reason of the failure is that Lorentzian KNTN geometry contains \emph{Dirac-Misner type singularity}, which is a manifestation 
of a non-trivial topological twist of the manifold. The BH \emph{mass formula} and \emph{Christodoulou-Ruffini mass formula} 
for  KNTN black holes are also derived. Finally, we compute the area bound  which is just Penrose like 
inequality for event horizon. From area bound we derive entropy bound. These thermodynamic products on the 
multi horizon playing a crucial role in BH thermodynamics to understand the microscopic nature of BH entropy.
\end{abstract}

%\ccode{PACS Nos.: include PACS Nos.}

\section{Introduction}
There has recently been investigation is focused on the area product and entropy product of the multi-horizon 
BHs in order to understand the BH entropy ${\cal S}=\frac{\cal A}{4}$ at the microscopic level. Previous studies 
suggested that the entropy product and area product for multi-horizon stationary, axisymmetric BHs are 
often(but not always) independent of the ADM(Arnowitt-Deser-Misner) mass and depends only on the quantized 
charges, quantized angular momentum etc. 
Firstly, in string theory and M-theory community, peoples  suggested that the entropy of the event horizon 
(${\mathcal H}^{+}$) and Cauchy horizons (${\mathcal H}^{-}$) are of the form \cite{mcdy96,mcfl97,mcflb97}:
${\cal S}_{\pm}=2\pi (\sqrt{N_{L}} \pm \sqrt{N_{R}})$, where the quantities $N_{L}$ and $N_{R}$ are the 
left and right moving  modes of a weakly coupled 2D conformal field theory(CFT). Therefore the product 
${\cal S}_{+}{\cal S}_{-}=4\pi^2(N_{L}-N_{R})$ should quantized in integer multiples of $4\pi^2$ 
\cite{mcfl97,mcflb97,mcgw11,pope14}. Indeed, one could find 
\begin{eqnarray}
\frac{{\cal S}_{+} {\cal S}_{-}}{(2\pi)^2} &=& J^2+Q_{1}Q_{2}Q_{3}Q_{4} ~.\label{CV4}
\end{eqnarray}
and
\begin{eqnarray}
\frac{{\cal S}_{+} {\cal S}_{-}}{(2\pi)^2} &=& J_{1}J_{2}+Q_{1}Q_{2}Q_{3} ~.\label{CV3}
\end{eqnarray}
for four and five dimensional BHs respectively. These expressions are modulus independent and strictly 
depends upon quantized charges and quantized angular momenta. In four dimension, these solutions are 
specified by four charges $Q_{1}, Q_{2}, Q_{3}, Q_{4}$ and one angular momentum $J$. Whereas in five 
dimension, they specified by three charges $Q_{1}, Q_{2}, Q_{3}$ and two angular momenta $J_{1}, J_{2}$ 
in addition to their mass. 

On the other hand, in general relativity community, peoples\cite{ansorg2,visser1,chen,pp14} 
pointed out that the entropy product of the event horizon and the Cauchy horizons for various 
stationary axially symmetric space-time  particularly in case of KN BH has the following form:
\begin{eqnarray}
\frac{{\cal S}_{+} {\cal S}_{-}}{(2\pi)^2} &=& J^2+\frac{Q^4}{4} ~.\label{KN1}
\end{eqnarray}
which is remarkably independent of the mass $({\cal M})$ parameter. But they depends on
the angular momentum $J$ and charge $Q$  of the BH  respectively.

The universal nature of BH entropy product rules for black ring and black string solutions 
have been investigated in \cite{castro}. Probably, Curir \cite{curir} in 1979 first calculated 
the area sum and entropy sum of Kerr BH for interpretation of the spin entropy of the area of 
the inner horizon. In \cite{wang}, the authors have examined a new universal property of 
entropy sum relation for various BHs in the asymptotically AdS space-time background.

The crucial point is that the entropy product of inner and outer horizons could be used to 
determine whether the Bekenstein-Hawking entropy may be written as a Cardy formula, therefore 
providing some evidence for a conformal field theory(CFT) description of the corresponding 
microstates\cite{castro,det}. Thus, we need to study inner horizon thermodynamics properly 
with outer horizon thermodynamics.

It is a well known fact that Cauchy horizon is an infinite blue-shift surface whereas event 
horizon is an infinite red-shift surface \cite{sch}. Thus when an observer cross the surface 
$r=r_{+}$, following a future directed time-like trajectory, is forever `lost' to an external 
observer and any radiation transmitted by such an observer at the moment of crossing will be 
\emph{infinitely red shifted}. Whereas, the same observer when cross the the surface $r=r_{-}$, 
following a future directed time-like trajectory and at the moment of crossing he/she will 
observe a panorama of the entire history of the external universe and any radiation transmitted 
by such an observer at the moment of crossing will be \emph{infinitely blue shifted}. This is 
the fundamental differences between  the ${\cal H}^+$ and ${\cal H}^-$.

The entropy and area product of multi-horizons are not always universal(mass-independence) and they 
sometimes fail also\cite{visser1}. We will show this properties are valid also for KNTN space-time \cite{miller,bini}.  
Due to the presence of gravito-magnetic mass or NUT parameter what would be the changes are manifested 
in the area product, entropy product, entropy sum, area sum, Smarr's mass formula \cite{smarr,smarr1} and 
Christodoulou-Ruffini mass formula in comparison with KN BH. This is the prime objective of this paper. We also verify explicitly that \emph{the first law of BH thermodynamics} does not hold for KNTN space-time. We further 
prove  that due to the presence of the NUT parameter Smarr-Gibbs-Duhem relation does not hold for KNTN space-time. 
It is also shown to be unlikely that such relations exist for KN BHs. 

Using this entropy product rules, we also derive the entropy bound, area bound and irreducible mass 
bound for both the horizons. These are basically all geometrical bound which was first proposed by 
Penrose\cite{peni} and what we now called it Penrose inequality. The upper bound of area of event 
horizon is exactly the Penrose inequality.

The structure of the paper is as follows.
In section (\ref{kntNUT}), we prove that the area product or entropy product,
area sum or entropy sum, the surface gravity product or surface temperature product or BH
temperature product of both the inner horizon and outer horizons do not shows any universal
properties due to the mass dependence. Such products seem to be \emph{not} universal in nature.
In section (\ref{smar}), we  explicitly show that
the BH mass or ADM mass can be expressed as in terms of the area  of both horizons ${\cal H}^{\pm}$. For our record,
we derive  in section (\ref{ruffini}), the Christodoulou-Ruffini \cite{cd,cr} mass formula for KNTN space-time. We 
also point out that the product of Christodoulou's irreducible mass of inner horizon (Cauchy horizon) and outer 
horizon (event horizon) are not \emph{independent} of mass. Finally we conclude in section (\ref{dis}).

\section{\label{kntNUT} Kerr-Newman-Taub-NUT Geometry:}

In Boyer-Lindquist like spherical coordinates $(t, r, \theta, \phi)$ the KNTN is completely determined by the  
four parameters i.e., the mass  $({\cal M})$,  charge ($Q$), angular momentum ($J=a{\cal M}$) and  gravito-magnetic 
monopole or NUT parameter ($ n$) or magnetic mass. Thus the corresponding metric \cite{miller,bini} is described by
\begin{eqnarray}
ds^2 &=& -\frac{\Delta}{\rho^2} \, \left[dt-P d\phi \right]^2+\frac{\sin^2\theta}{\rho^2} \, \left[(r^2+a^2+n^2) 
\,d\phi-a dt\right]^2 +\rho^2 \, \left[\frac{dr^2}{\Delta}+d\theta^2\right] ~.\label{nkntn}
\end{eqnarray}

where
\begin{eqnarray}
a &\equiv&\frac{J}{{\cal M}},\, \rho^2 \equiv r^2+(n+a\cos\theta)^2 \\
\Delta &\equiv& r^2-2{\cal M}r+a^2+Q^2-n^2\\
P &\equiv& a\sin^2\theta-2n\cos\theta  ~.\label{adeltap}
\end{eqnarray}
The electromagnetic field 2-form would be given by
$$
F = \frac{Q}{\rho^4}[r^2-(n+a\cos\theta)^2]dr\wedge (dt-P d\phi) +
$$
\begin{eqnarray}
\frac{2aQr\sin\theta\cos\theta}{\rho^4}
d\theta \wedge[(r^2+a^2+n^2) \,d\phi-a dt] ~.\label{fieldt}
\end{eqnarray}

Note that when $Q=0$, the electromagnetic field tensor vanishes and the metric satisfies the vacuum Einstein equations. When $n=0$, the specific geometry reduces to Kerr-Newman geometry and when $Q=n=0$, the geometry reduces to Kerr geometry.

The remarkable feature of this space-time is that:  it has duality between the mass and the NUT parameter. 
Therefore the solution is invariant under the duality transformation, 
${\cal M} \leftrightarrow i n$, $r \leftrightarrow i a \chi $, where $\chi$ is an angle coordinate 
and $a$ is the Kerr parameter. This is in fact duality between gravito-electric mass $({\cal M})$ 
and gravito-magnetic charges ($n$) \cite{dad}.

The radius of the horizon is determined by the solution of the function
$\Delta=0$.
i.e.,
\begin{eqnarray}
r=r_{\pm}\equiv {\cal M}\pm\sqrt{{\cal M}^2-a^2-Q^2+n^2}\,\, \mbox{and}\,\,  r_{+}> r_{-}
\end{eqnarray}

The static limit surface (outer region of the ergo-sphere)  is  at $g_{tt}=0$ i.e.
\begin{eqnarray}
r=r_{ergo}\equiv {\cal M}+\sqrt{{\cal M}^2-a^2\cos^2\theta-Q^2+n^2}\,\, ~.\label{ergokntn}
\end{eqnarray}
As long as
\begin{eqnarray}
{\cal M}^2+n^2 - Q^2 - a^2 \geq 0 ~.\label{ineq}
\end{eqnarray}
then the KNTN metric  describes a BH, otherwise it has a naked ringlike singularity.
When ${\cal M}^2+n^2 -Q^2 - a^2=0$, the situation is called extremal situation in
gravitational physics.

Now the area of both the horizons (${\cal H}^\pm$) for KNTN space-time is
\begin{eqnarray}
{\cal A}_{\pm} &=& \int^{2\pi}_0\int^\pi_0  \sqrt{g_{\theta\theta}g_{\phi\phi}} d\theta d\phi =4\pi(r_{\pm}^2+a^2+n^2) \\
                &=& 4\pi\left[2({\cal M}^2+n^2)-Q^2 \pm 2{\cal M}\sqrt{{\cal M}^2+n^2-a^2-Q^2} \right]
~.\label{arKNTN}
\end{eqnarray}

The angular velocity of ${\cal H}^\pm$ is
\begin{eqnarray}
 {\Omega}_{\pm} &=& \frac{a}{r_{\pm}^2+a^2+n^2} =\frac{a}{\left(2{\cal M}r_{\pm}-Q^2+2n^2 \right)} ~. \label{omkntn}
\end{eqnarray}

The semi classical Bekenstein-Hawking \cite{Beken,bcw} entropy of ${\cal H}^\pm$ reads
(in units in which $G=\hbar=c=1$)
\begin{eqnarray}
{\cal S}_{\pm} &=& \frac{{\cal A}_{\pm}}{4} =\pi(r_{\pm}^2+a^2+n^2)=\pi \left(2{\cal M}r_{\pm}-Q^2+2n^2 \right)  ~.\label{etpKntn}
\end{eqnarray}

The surface gravity of ${\cal H}^\pm$ is
\begin{eqnarray}
{\kappa}_{\pm} &=& \frac{r_{\pm}-r_{\mp}}{2(r_{\pm}^2+a^2+n^2)}= \frac{r_{\pm}-r_{\mp}}
{ 2\left(2{\cal M}r_{\pm}-Q^2+2n^2 \right)} \,\, \mbox{and}\,\,  \kappa_{+}> \kappa_{-} ~.\label{sgKNtn}
\end{eqnarray}
and
the BH temperature or Hawking temperature of ${\cal H}^\pm$ reads as
\begin{eqnarray}
T_{\pm} &=& \frac{{\kappa}_{\pm}}{2\pi} =\frac{r_{\pm}-r_{\mp}}{4\pi (r_{\pm}^2+a^2+n^2)}  ~.\label{tmKNtn}
\end{eqnarray}
It should be noted that $T_{+} > T_{-}$.

Finally, the horizon Killing vector field may be defined for ${\cal H}^\pm$ is
\begin{eqnarray}
{\chi_{\pm}}^{a} &=& (\partial_{t})^a +\Omega_{\pm} (\partial_{\phi})^a~.\label{hkv}
\end{eqnarray}
Their are some useful relations noted as:
\begin{eqnarray}
r_{+} r_{-} &=& a^2+Q^2 -n^2 \\
r_{+} + r_{-} &=& 2 {\cal M} ~.\label{rKNtn}
\end{eqnarray}

Remarkably, if the inner Cauchy horizon exists (i.e. if $J$, $Q$ and $n$
do not vanish simultaneously ), then the product of the area of the
Cauchy horizon and event horizon  of  KNTN space-time holds the following
relation:
\begin{eqnarray}
{\cal A}_{+} {\cal A}_{-} &=& (8\pi)^2\left(J^2+\frac{Q^4}{4}+
n^2({\cal M}^2+n^2-Q^2)\right) ~.\label{proarKNtn}
\end{eqnarray}
It seems that the area product strictly depends upon the mass of the
space-time. Thus this product is not universal in this sense for stationary, axially
symmetric KNTN space-time.

Now the sum of the area of both the horizons are
\begin{eqnarray}
{\cal A}_{+}+ {\cal A}_{-} &=& 8\pi\left[2({\cal M}^2+n^2)-Q^2)\right] ~.\label{sumKNtn}
\end{eqnarray}
It shows that the area sum of both the horizons also depend on the mass, charge and the NUT parameter.
It does not manifested of any universal character on the ``area sum''.

Also the product of entropy is
\begin{eqnarray}
{\cal S}_{+} {\cal S}_{-} &=& (2\pi)^2\left(J^2+\frac{Q^4}{4}+n^2({\cal M}^2+n^2-Q^2)\right) ~.\label{pentkntn}
\end{eqnarray}
It is also depends on  the mass $({\cal M})$ parameter.
Similarly, for our completeness, we also calculate  the ``entropy sum'' of
both the horizons and found to be
\begin{eqnarray}
{\cal S}_{+}+ {\cal S}_{-} &=& 2\pi \left[2({\cal M}^2+n^2)-Q^2\right] ~.\label{entrokntn}
\end{eqnarray}
For our record, we also compute 
the entropy minus of ${\cal H}^\pm$ as 
\begin{eqnarray}
{\cal S}_{\pm}- {\cal S}_{\mp} &=& 8\pi {\cal M} T_{\pm}{\cal S}_{\pm}  ~.\label{entr-}
\end{eqnarray}
and  
the entropy division of ${\cal H}^\pm$ as
\begin{eqnarray}
\frac{{\cal S}_{+}}{{\cal S}_{-}} &=& - \frac{T_{-}}{T_{+}}=\frac{\Omega_{-}}{\Omega_{+}} ~.\label{entrd}
\end{eqnarray}
Again, the sum of entropy inverse is found to be 
\begin{eqnarray}
\frac{1}{{\cal S}_{+}}+\frac{1}{{\cal S}_{-}} &=& \frac{2({\cal M}^2+n^2)-Q^2}
{2 \pi\left(J^2+\frac{Q^4}{4}+n^2({\cal M}^2+n^2-Q^2)\right)} ~.\label{etpid}
\end{eqnarray}
and the minus of entropy inverse is computed to be 
\begin{eqnarray}
\frac{1}{{\cal S}_{\pm}}-\frac{1}{{\cal S}_{\mp}} &=&\mp \frac{{\cal M}\sqrt{{\cal M}^2+n^2-a^2-Q^2}}
{2\pi\left(J^2+\frac{Q^4}{4}+n^2({\cal M}^2+n^2-Q^2)\right)} ~.\label{etpid-}
\end{eqnarray}
It indicates that they all are mass dependent relations. There are some useful relations for 
KNTN BH as derived by Curir\cite{curir} for Kerr BH:
\begin{eqnarray}
T_{+}{\cal S}_{+}+ T_{-}{\cal S}_{-} &=& 0 ~.\label{exxkn1}
\end{eqnarray}
and
\begin{eqnarray}
\frac{{\Omega}_{+}}{T_{+}}+\frac{{\Omega}_{-}}{T_{-}} &=& 0 ~. \label{omknn}
\end{eqnarray}
The above important thermodynamic products of multi horizons could be used to determine
the classical BH entropy in terms of Cardy formula, therefore giving some evidence for a 
BH/CFT description of the corresponding microstates\cite{castro}. It has been also shown that
from the above Eq. \ref{exxkn1} the central charge being the same for 2-horizon BHs.

Based on these above thermodynamic relations, we are now able to compute the entropy bound  
for KNTN BH. From the Eq. \ref{ineq}, we obtain Kerr like bound for KNTN BH:
\begin{eqnarray}
{\cal M}^4+(n^2- Q^2){\cal M}^2 - J^2 \geq 0 ~.\label{ineq1}
\end{eqnarray}
or
\begin{eqnarray}
{\cal M}^2 &=& \frac{-(n^2- Q^2)+\sqrt{(n^2- Q^2)^2+4J^2}}{2} ~.\label{ieq}
\end{eqnarray}
Since $r_{+} \geq r_{-}$, one obtains ${\cal S}_{+} \geq {\cal S}_{-} \geq 0$.
Then the entropy product (\ref{pentkntn}) gives:
\begin{eqnarray}
{\cal S}_{+} \geq  \sqrt{{\cal S}_{+} {\cal S}_{-}}=\sqrt{(2\pi)^2\left(J^2+\frac{Q^4}{4}+n^2({\cal M}^2+n^2-Q^2) \right)}
\geq {\cal S}_{-}
~.\label{ieq1}
\end{eqnarray}
and the entropy sum gives:
\begin{eqnarray}
2\pi \left[2({\cal M}^2+n^2)-Q^2\right] = {\cal S}_{+}+ {\cal S}_{-} \geq {\cal S}_{+}
\geq \frac{{\cal S}_{+}+ {\cal S}_{-}}{2}= \pi \left[2({\cal M}^2+n^2)-Q^2\right] \geq {\cal S}_{-}  ~.\label{inq2}
\end{eqnarray}
Thus the entropy bound for  ${\cal H}^{+}$:
\begin{eqnarray}
 \pi \left[2({\cal M}^2+n^2)-Q^2\right]  \leq {\cal S}_{+} \leq 2\pi \left[2({\cal M}^2+n^2)-Q^2\right]   ~.\label{inq3}
\end{eqnarray}
and  the entropy bound for  ${\cal H}^{-}$:
\begin{eqnarray}
 0 \leq {\cal S}_{-} \leq   \sqrt{(2\pi)^2\left(J^2+\frac{Q^4}{4}+n^2({\cal M}^2+n^2-Q^2) \right)} ~.\label{inq4}
\end{eqnarray}
From this bound one can get area bound which can be found in the latter section. Thus the universal behavior 
of area product, entropy product, area sum and entropy sum do not hold  for KNTN space-time. It should be noted 
that in the limit $Q=0$, we obtain the entropy bound for KTN BH \cite{xu}.

In appendix A, we have computed various thermodynamic quantities for KN BH, Kerr BH in comparison
with KNTN BH.

\section{\label{smar} The Mass Formula for  KNTN Spacetime:}

To evaluate the mass formula for KNTN BH, first we need to compute the surface area of this BH. The surface 
area of the BH is the
two dimensional surface formed by the intersection of the outer horizon or event horizon with a space-like 
hyper-surface. According to Hawking's area theorem \cite{bcw}, the surface area of a BH always increases i.e.,
\begin{eqnarray}
d{\cal A} \geq 0 \label{areahaw}
\end{eqnarray}
Now for KNTN BH the surface area  of both event horizon and Cauchy horizon is indeed constant.  can be defined as
\begin{eqnarray}
{\cal A}_{\pm} &=& 4\pi \left( 2{\cal M}^2-Q^2 +2n^2 \pm 2\sqrt{{\cal M}^4-J^2-{\cal M}^2Q^2+n^2{\cal M}^2}\right) ~.\label{arKNtn}
\end{eqnarray}
It may be noted that ${\cal A}_{+} >{\cal A}_{-}$.

Inverting the above relation one can obtain the BH mass or ADM mass can be
expressed as in terms of area of both the horizons ${\cal H}^{\pm}$,
\begin{eqnarray}
{\cal M}^2 &=&  \frac{1}{({\cal A}_{\pm}-4\pi n^2)} \left[ \frac{{\cal A}_{\pm}^2}{16\pi}+4\pi J^2
+\frac{{\cal A}_{\pm}Q^2}{2}+\pi Q^4 -n^2({\cal A}_{\pm}-4\pi n^2-4\pi Q^2)\right]~.\label{masskntn}
\end{eqnarray}
It is remarkable that the mass can be expressed as in terms of both the area of
${\cal H}^+$ and ${\cal H}^-$.

Now we can deduce  the mass differential for KNTN space-time. It is indeed expressed as four
physical invariants of both ${\cal H}^+$ and ${\cal H}^-$,
\begin{eqnarray}
d{\cal M} &=& \Gamma_{\pm} d{\cal A}_{\pm} + \Omega_{\pm} dJ +\Phi_{\pm}dQ+\Phi_{\pm}^{n}dn
~. \label{dm}
\end{eqnarray}
where
\begin{eqnarray}
\Gamma_{\pm} &=&  \frac{1}{2{\cal M}({\cal A}_{\pm}-4\pi n^2)^2} \left( \frac{{\cal A}_{\pm}^2}{16\pi}-4\pi J^2-\pi Q^4
-\frac{n^2{\cal A}_{\pm}}{2} +2\pi n^2Q^2   \right) \\
\Omega_{\pm} &=& \frac{4\pi J}{{\cal M}{(\cal A}_{\pm}-4\pi n^2)} \\
\Phi_{\pm}^{Q} &=& \frac{1}{2{\cal M}({\cal A}_{\pm}-4\pi n^2)} \left( Q{\cal A}_{\pm}+4\pi Q^3-8\pi n^2Q \right)\\
\Phi_{\pm}^n &=& \frac{\left(16\pi n^3 {\cal A}_{\pm}-4n\pi Q^2 {\cal A}_{\pm}-\frac{3}{2}n {{\cal A}_{\pm}}^2
-32\pi^2 n^5+32n\pi J^2+8n \pi^2Q^4 \right)}{2{\cal M}({\cal A}_{\pm}-4\pi n^2)^2}
~. \label{invkntn}
\end{eqnarray}
where
\begin{eqnarray}
 \Gamma_{\pm} &=& \mbox{Effective surface tension of ${\cal H}^{+}$ and ${\cal H}^{-}$} \nonumber \\
\Omega_{\pm} &=&  \mbox{Angular velocity of ${\cal H}^\pm$} \nonumber \\
\Phi_{\pm}^{Q} &=& \mbox{Electromagnetic potentials of ${\cal H}^\pm$ for electric charge} \\
\Phi_{\pm}^n &=& \mbox{Electromagnetic potentials of
${\cal H}^\pm$ for NUT charge} \nonumber
\end{eqnarray}

Using the Euler's theorem on homogenous functions to ${\cal M}$ of degree $\frac{1}{2}$ in $({\cal A}_{\pm}, J, Q^2, n^2)$. 
Thus  we can deduce the mass can be expressed in terms of these quantities both for ${\cal H}^\pm$ as a  bilinear form
\begin{eqnarray}
{\cal M} &=& 2\Gamma_{\pm} {\cal A}_{\pm} + 2J \Omega_{\pm} + \Phi_{\pm}^{Q} Q +\Phi_{\pm}^{n} n
~. \label{bilinear}
\end{eqnarray}
This has been derived from the homogenous function of degree $\frac{1}{2}$ in
$({\cal A}_{\pm}, J, Q^2, n^2)$.

Remarkably, $\Gamma_{\pm}$,  $\Omega_{\pm}$, $\Phi_{\pm}^{Q}$ and $\Phi_{\pm}^n$ can be defined and are 
constant on the ${\cal H}^+$ and ${\cal H}^-$ for any stationary, axially symmetric space-time.

Since the $d{\cal M}$ is perfect differential, one may choose freely any path of
integration in $({\cal A}_{\pm}, J, Q, n)$ space. Thus the surface energy
${\cal E}_{s \pm}$ for ${\cal H}^+$ and ${\cal H}^{-}$ can be defined as

\begin{eqnarray}
{\cal E}_{s \pm} &=& \int_{0}^{{\cal A}_{\pm}} \Gamma (\tilde{{\cal A}_{\pm}}
, 0 ,0,0) d\tilde{{\cal A}_{\pm}}; ~ \label{se}
\end{eqnarray}

The rotational energy  for ${\cal H}^+$ and ${\cal H}^{-}$ can be defined by
\begin{eqnarray}
{\cal E}_{r \pm} &=& \int_{0}^{J} \Omega_{\pm} ({\cal A}_{\pm}
, \tilde{J} , 0, 0) d\tilde{J},\,\,  \mbox{${\cal A}_{\pm}$ fixed}; ~ \label{re}
\end{eqnarray}

The electromagnetic energy  for ${\cal H}^+$ and ${\cal H}^{-}$ due to the charge $Q$
is given by
\begin{eqnarray}
{\cal E}_{em \pm} &=& \int_{0}^{Q} \Phi_{\pm} ({\cal A}_{\pm}
, J, \tilde{Q}, 0) d\tilde{Q},\,\, \mbox{${\cal A}_{\pm}$, $J$  fixed}; ~ \label{re1}
\end{eqnarray}
and finally  the electromagnetic  energy for ${\cal H}^+$ and ${\cal H}^{-}$ due to the NUT charge $n$ is given by
\begin{eqnarray}
{\cal E}_{em \pm} &=& \int_{0}^{n} \Phi_{\pm} ({\cal A}_{\pm}
, J, \tilde{Q}, \tilde{n}) d\tilde{n},\,\, \mbox{${\cal A}_{\pm}$, $J$, $\tilde{Q}$ fixed}; ~ \label{re2}
\end{eqnarray}

Generally, combining the mass differential Eq. (\ref{dm}) with the first law leads to the Smarr-Gibbs-Duhem relation. We observe
that from Eqs. (\ref{dm}) and (\ref{bilinear}), such an equation does not hold for KNTN space-time because
\begin{eqnarray}
\Gamma_{\pm} &=& \frac{\partial {\cal M}}{\partial {\cal A}_{\pm}} \neq \frac{{\kappa}_{\pm}} {8\pi}=\frac{T_{\pm}}{4}
\end{eqnarray}
An important point should be noted here that \emph{the first law of BH thermodynamics} do not satisfied for KNTN space-time.
The above features do not hold because the Lorentzian KNTN space-time containing Dirac-Misner type singularities, a 
coordinate singularity which may be considered as a manifestation of a non-trivial topological twisting
of the manifold\cite{gibbons}.

\section{\label{ruffini} Christodoulou's Irreducible Mass bound for KNTN Space-time:}

Floyd and Penrose  were first noticed that in an example of the extraction of the energy from a Kerr BH,
 the surface area of the horizon increases when a BH undergoing  any transformations. What is now called the
Penrose process \cite{penrose}. Independently, Christodoulou \cite{cd} had shown that a quantity which he named
the ``irreducible mass '' of the BH, $ {\cal M}_{irr}$ unchanged by means of any process. In fact , most
processes result in an increase in  $ {\cal M}_{irr}$  and during reversible process this quantity also does
not change. This result indicates that there exist a relation between area and irreducible mass. Now it
is well known that  $ {\cal M}_{irr}$ is proportional to the square root of the BH's area.  Since the KNTN
space-time  has regular event horizon and Cauchy horizons . Thus the irreducible mass can be defined
for both the horizons are
\begin{eqnarray}
 {\cal M}_{irr, \pm} &=& \sqrt{\frac{{\cal A}_{\pm}}{16\pi}} =\frac{\sqrt{r_{\pm}^2+a^2+n^2}}{2}
~. \label{irrm}
\end{eqnarray}
where $+$ indicates for ${\cal H}^+$ and $-$ indicates for ${\cal H}^-$.
The area and angular velocity can be expressed as in terms of ${\cal M}_{irr, \pm}$:
\begin{eqnarray}
{\cal A}_{\pm} &=& 16 \pi ({\cal M}_{irr, \pm})^2 ~. \label{irrma}
\end{eqnarray}
and
\begin{eqnarray}
 {\Omega}_{\pm} &=& \frac{a}{r_{\pm}^2+a^2+n^2} = \frac{a}{4({\cal M}_{irr, \pm})^2} ~. \label{iromega}
\end{eqnarray}

Interestingly, the product of the irreducible mass of  ${\cal H}^\pm$  equivalent to area product for KNTN
space-time:

\begin{eqnarray}
 {\cal M}_{irr,+} {\cal M}_{irr,-} &=& \sqrt{\frac{{\cal A}_{+} {\cal A}_{-}}{(16\pi)^{2}}} \\
                                 &=&  \sqrt{J^2+\frac{Q^4}{4}+n^2({\cal M}^2+n^2-Q^2)} ~. \label{irrmp}
\end{eqnarray}
It shows that this product is not \emph {universal} because it depends on the
mass parameter.

In fact the Christodoulou-Ruffini mass formula for KNTN space-time in term of  irreducible mass ${\cal M}_{irr, \pm}$,  
angular momentum $J$ , charge $Q$ and NUT parameter $n$ is given by
$$
{\cal M}^2 = \left[({\cal M}_{irr, \pm}+\frac{Q^2}{4 {\cal M}_{irr, \pm} })^2+
\frac{J^2}{4 ({\cal M}_{irr, \pm})^2}-n^2\left( 1+\frac{Q^2}{4({\cal M}_{irr, \pm})^2}-
\frac{n^2}{4 ({\cal M}_{irr, \pm})^2}\right) \right] \times
$$
\begin{eqnarray}
 \left( 1-\frac{n^2}{4 ({\cal M}_{irr, \pm})^2}\right)^{-1}  ~. \label{irrma1}
\end{eqnarray}
When the NUT parameter and charge parameter goes to zero we get the mass formula for Kerr-Newman 
space-time \cite{cr}. When the charge parameter vanishes, we obtain the mass formula for Kerr-Taub-NUT 
space-time:
\begin{eqnarray}
{\cal M}^2 = \left[({\cal M}_{irr, \pm})^2+\frac{J^2}{4 ({\cal M}_{irr, \pm})^2}-n^2\left( 1-\frac{n^2}
{4 ({\cal M}_{irr, \pm})^2}\right) \right] \left( 1-\frac{n^2}{4 ({\cal M}_{irr, \pm})^2}\right)^{-1}  ~. \label{irrma2}
\end{eqnarray}
Again when $n=0$, we recover the mass formula for Kerr BH \cite{cr}. 

Now we would like to compute the area bound and irreducible mass bound for KNTN BH followed by the previous section. 
Since $r_{+} \geq r_{-}$, one obtains ${\cal A}_{+} \geq {\cal A}_{-} \geq 0$.
Then the area product gives:
\begin{eqnarray}
{\cal A}_{+}  \geq  \sqrt{{\cal A}_{+} {\cal A}_{-}}=\sqrt{(8\pi)^2\left(J^2+\frac{Q^4}{4}+n^2({\cal M}^2+n^2-Q^2)\right)}
\geq {\cal A}_{-}
~.\label{ieqa1}
\end{eqnarray}
and the area sum gives:
\begin{eqnarray}
8\pi \left[2({\cal M}^2+n^2)-Q^2\right] = {\cal A}_{+}+ {\cal A}_{-} \geq {\cal A}_{+}
\geq \frac{{\cal A}_{+}+ {\cal A}_{-}}{2}= 4\pi \left[2({\cal M}^2+n^2)-Q^2\right] \geq {\cal A}_{-} ~.\label{inqa2}
\end{eqnarray}
Thus the area bound for  ${\cal H}^{+}$:
\begin{eqnarray}
 4\pi \left[2({\cal M}^2+n^2)-Q^2\right]  \leq {\cal A}_{+} \leq 8\pi \left[2({\cal M}^2+n^2)-Q^2\right]   ~.\label{inqa3}
\end{eqnarray}
and  the area bound for  ${\cal H}^{-}$:
\begin{eqnarray}
 0 \leq {\cal A}_{-} \leq  \sqrt{(2\pi)^2\left(J^2+\frac{Q^4}{4}+n^2({\cal M}^2+n^2-Q^2) \right)} ~.\label{inqa4}
\end{eqnarray}
From this area bound, we get irreducible mass bound for KNTN BH:
For ${\cal H}^{+}$:
\begin{eqnarray}
\frac{\sqrt{\left[2({\cal M}^2+n^2)-Q^2\right]}}{2} \leq {\cal M}_{irr, +} \leq 
\frac{\sqrt{\left[2({\cal M}^2+n^2)-Q^2\right]}}{\sqrt{2}}   ~.\label{inqa5}
\end{eqnarray}
and  for ${\cal H}^{-}$:
\begin{eqnarray}
0 \leq {\cal M}_{irr,-} \leq  \frac{\left(J^2+\frac{Q^4}{4}+n^2({\cal M}^2+n^2-Q^2)\right)^\frac{1}{4}}
{\sqrt{2}} ~.\label{inqa6}
\end{eqnarray}
Eq. \ref{inqa5} is nothing but the Penrose inequality, which is the first geometric inequality for 
BHs\cite{peni}.

So far we derived different thermodynamic formulae and these formulae is expected to be 
useful to further understanding the microscopic nature of BH entropy both exterior and 
interior.

\section{\label{dis} Discussion:}
In this work, we have studied the universal nature of the product and sum of the entropy(area) 
of the event horizon and Cauchy horizons for KNTN BH. We also examined entropy minus and entropy 
division of both the horizons. We found that all the thermodynamic relations are mass dependent.
Thus the universal nature of the above features are fail in KNTN geometry.

Based on these relations, we also derived the entropy bound and area bound
for all the horizons. Furthermore, we have derived the Smarr\cite{smarr} formula 
and Christodoulou-Ruffini mass formula for KNTN space-time. Moreover, we calculate the 
irreducible mass bound for this type of BH. These formulae might have useful to understanding 
the microscopic nature of BH entropy both exterior and interior. Again, the entropy products 
of inner horizon and outer horizons could be used to determine whether the classical BH entropy 
could be written as a Cardy formula, giving some evidence for a holographic description of BH/CFT 
correspondence\cite{chen}. The above thermodynamic properties including the Hawking temperature 
and area of both the horizons may therefore be expected to play a crucial role in understanding the 
BH entropy at the microscopic level.

The failure of the first law of BH thermodynamics and  Smarr-Gibbs-Duhem  relations  for the  
KNTN spacetime due to the presence of the non-trivial NUT parameter.  Such results are unlikely in 
the KN BH. It is  known that in the presence of NUT charges, the entropy of a space-time acquires a
contribution from the Dirac-Misner strings \cite{gibbons} that the NUT charge introduces: this is in addition 
to the usual contribution from horizon area . Thus it seems quite natural to expect that the formula for the 
area product, entropy product, entropy sum and area sum get modified due to the Dirac-Misner strings. This 
investigation might also be interesting and it could be found elsewhere.

%\appendix
\section{Appendix:}
\begin{center}
\begin{tabular}{|c|c|c|c|}
    \hline
    % after \\: \hline or \cline{col1-col2} \cline{col3-col4} ...
    Parameter & KN BH & Kerr BH  & KNTN BH\\
    \hline
    $r_{\pm}$: &${\cal M}\pm\sqrt{{\cal M}^{2}-a^2-Q^{2}}$ &$M\pm\sqrt{{\cal M}^{2}-a^2}$ &
    ${\cal M}\pm\sqrt{{\cal M}^{2}+n^2-a^2-Q^{2}} $\\

    $\sum r_{i}$: & $2{\cal M}$ & $2{\cal M}$ & $2{\cal M}$\\

    $\prod r_{i}$: & $a^2+Q^2$ & $a^2$ & $a^2+Q^2-n^2$ \\

    ${\cal A}_{\pm}$: & $4\pi \left(2{\cal M}r_{\pm}-Q^2\right)$ & $8\pi {\cal M}r_{\pm}$ &
    $4\pi \left(2{\cal M}r_{\pm}-Q^2+2n^2 \right)$ \\

    $\sum {\cal A}_{i}$ : & $8\pi\left(2{\cal M}^2-Q^2\right)$ & $16\pi {\cal M}^2$ &
    $8\pi\left(2{\cal M}^2+2n^2-Q^2\right)$\\

    $\prod {\cal A}_{i}$: &$(8\pi)^2\left(J^2+\frac{Q^4}{4}\right)$ & $\left(8\pi J\right)^2$
    & $(8\pi)^2\left[J^2+\frac{Q^4}{4}+n^2({\cal M}^2+n^2-Q^2)\right]$ \\

    $S_{\pm}$: &$\pi \left(2{\cal M}r_{\pm}-Q^2\right) $ & $2\pi Mr_{\pm}$ &
    $\pi \left(2{\cal M}r_{\pm}-Q^2+2n^2 \right)$\\

    $\sum {\cal S}_{i}$ : & $2\pi(2{\cal M}^2-Q^2)$& $4\pi {\cal M}^2$ &
    $2\pi\left(2{\cal M}^2+2n^2-Q^2\right)$\\

    $\prod {\cal S}_{i}$: & $(2\pi)^2\left(J^2+\frac{Q^4}{4}\right)$ & $\left(2\pi J\right)^2$ &
    $(2\pi)^2\left[J^2+\frac{Q^4}{4}+n^2({\cal M}^2+n^2-Q^2)\right]$\\

    $\kappa_{\pm}$: &$ \frac{r_{\pm}-r_{\mp}}{2 \left(2{\cal M}r_{\pm}-Q^2 \right)}$ & $ \frac{r_{\pm}-r_{\mp}}{4{\cal M}r_{\pm}}$ &  $\frac{r_{\pm}-r_{\mp}}{2 \left(2{\cal M}r_{\pm}-Q^2+2n^2\right)}$\\

    $\sum{\kappa}_{i}$: &$\frac{4{\cal M}(a^2+Q^2-{\cal M}^2)}{(4J^2+Q^4)}$ & $\frac{a^2-{\cal M}^2}{aJ}$ &
    $\frac{{\cal M}(a^2+Q^2-{\cal M}^2-n^2)}{[J^2+\frac{Q^4}{4}+n^2({\cal M}^2+n^2-Q^2)]}$\\

    $\prod {\kappa}_{i}$: &$\frac{a^2+Q^2-{\cal M}^2}{(4J^2+Q^4)}$ & $\frac{a^2-{\cal M}^2}{4J^2}$ &
    $ \frac{(a^2+Q^2-{\cal M}^2-n^2)}{4[J^2+\frac{Q^4}{4}+n^2({\cal M}^2+n^2-Q^2)]}$\\

    $T_{\pm}$:   &$\frac{r_{\pm}-r_{\mp}}{4\pi (r_{\pm}^2+a^2)}$ & $\frac{r_{\pm}-r_{\mp}}{4\pi (r_{\pm}^2+a^2)}$ &
     $ \frac{r_{\pm}-r_{\mp}}{4\pi (r_{\pm}^2+a^2+n^2)}$\\

    $\sum T_{i}$:  &$\frac{2{\cal M}(a^2+Q^2-{\cal M}^2)}{\pi(4J^2+Q^4)} $ & $\frac{a^2-{\cal M}^2}{2\pi aJ}$ &
    $ \frac{{\cal M}(a^2+Q^2-{\cal M}^2-n^2)}{2 \pi [J^2+\frac{Q^4}{4}+n^2({\cal M}^2+n^2-Q^2)]}$\\

    $\prod T_{i}$ : &$\frac{(a^2+Q^2-{\cal M}^2)}{4 \pi^{2}(4J^2+Q^4)}$ & $\frac{a^2-{\cal M}^2}{(4\pi J)^2}$
    & $\frac{(a^2+Q^2-{\cal M}^2-n^2)}{(4\pi)^{2}[J^2+\frac{Q^4}{4}+n^2({\cal M}^2+n^2-Q^2)]} $\\

    ${\cal M}_{irr, \pm}$: &$\sqrt{\frac{2{\cal M}r_{\pm}-Q^2}{4}}$&$\sqrt{\frac{{\cal M}r_{\pm}}{2}}$ &
    $\sqrt{\frac{2{\cal M}r_{\pm}-Q^2+2n^2}{4}}$\\

    $\sum {\cal M}_{irr}^{2}$: &${\cal M}^2-\frac{Q^{2}}{2} $ & ${\cal M}^{2}$ & ${\cal M}^2+n^2-\frac{Q^{2}}{2}$\\

    $\prod {\cal M}_{irr}$: &$\sqrt{\frac{J^2+\frac{Q^4}{4}}{4}}$ & $\frac{J}{2}$ &
    $\sqrt{\frac{J^2+\frac{Q^4}{4}+n^2({\cal M}^2+n^2-Q^2)}{4}} $\\

    $\Omega_{\pm}$: &$\frac{a}{2{\cal M}r_{\pm}-Q^2}$ & $\frac{a}{2{\cal M}r_{\pm}}$ &
    $\frac{a}{2{\cal M}r_{\pm}+2n^2-Q^2}$ \\

    $\sum \Omega_{i} $: &$\frac{2a(2{\cal M}^2-Q^2)}{4J^2+Q^{4}}$ & $\frac{1}{a}$ &
    $\frac{a(2{\cal M}^2+2n^2-Q^2)}{2[J^2+\frac{Q^4}{4}+n^2({\cal M}^2+n^2-Q^2)]} $\\

    $\prod \Omega_{i}$ : &$\frac{a^{2}}{4J^2+Q^{4}}$ & $\frac{1}{4{\cal M}^{2}}$ &
    $\frac{a^2}{4[J^2+\frac{Q^4}{4}+n^2({\cal M}^2+n^2-Q^2)]}$ \\

\hline
\end{tabular}
\end{center}

\bibliographystyle{elsarticle-num}

\end{document}